\documentclass[%
superscriptaddress,
preprint,
 amsmath,amssymb,
 aps,
]{revtex4-1}

\usepackage{graphicx}
\usepackage{dcolumn}
\usepackage{bm}
\usepackage{xcolor}
\usepackage{qcircuit}
\usepackage{braket}
\usepackage{float}
\usepackage{algorithm}
\usepackage{algorithmicx}
\usepackage{algpseudocode}
\usepackage{mathtools}
\usepackage{hyperref}
\usepackage[normalem]{ulem}

\renewcommand{\figureautorefname}{Figure~\negthinspace}
\renewcommand{\equationautorefname}{Equation~\negthinspace}

\renewcommand{\sectionautorefname}{Section~\negthinspace}
\renewcommand{\appendixautorefname}{Appendix~\negthinspace}

\begin{document}

\preprint{BNL-NTU}

\title{Variational Quantum Reinforcement Learning via Evolutionary Optimization}

\author{Samuel Yen-Chi Chen}
\email{ychen@bnl.gov}
\affiliation{%
 Computational Science Initiative, Brookhaven National Laboratory, Upton, NY 11973, USA
}%
\author{Chih-Min Huang}
\email{b06501134@ntu.edu.tw}
\affiliation{%
 Department of Physics, National Taiwan University, Taipei 10617, Taiwan}%
\author{Chia-Wei Hsing}
\email{cwhsing0219@gmail.com}
\affiliation{%
 Department of Physics, National Taiwan University, Taipei 10617, Taiwan}%
\author{Hsi-Sheng Goan}%
 \email{goan@phys.ntu.edu.tw}
\affiliation{Department of Physics, National Taiwan University, Taipei 10617, Taiwan}
\affiliation{Center for Theoretical Physics, National Taiwan University, Taipei 10617, Taiwan}
\affiliation{Center for Quantum Science and Engineering, National Taiwan University, Taipei 10617, Taiwan}%
\affiliation{Physics Division, National Center for Theoretical Science, Taipei 10617, Taiwan}
\author{Ying-Jer Kao}
\email{yjkao@phys.ntu.edu.tw}
\affiliation{%
 Department of Physics, National Taiwan University, Taipei 10617, Taiwan}%
 \affiliation{Center for Theoretical Physics, National Taiwan University, Taipei 10617, Taiwan}
\affiliation{Center for Quantum Science and Engineering, National Taiwan University, Taipei 10617, Taiwan}%



\date{\today}

\begin{abstract}

Recent advance in classical reinforcement learning (RL) and quantum computation (QC) points to a promising direction of performing RL on a quantum computer.
However, potential applications in quantum RL are limited by the  number of qubits available in the modern quantum devices. 
%
Here we present two frameworks of deep quantum RL tasks using a gradient-free evolution optimization: First, we apply the amplitude encoding scheme to the Cart-Pole problem; Second,  we propose  a hybrid framework where the quantum RL agents are equipped with hybrid tensor network-variational quantum circuit (TN-VQC) architecture to handle inputs with dimensions exceeding the number of qubits. This allows us to perform quantum RL on the MiniGrid environment with 147-dimensional inputs. 
 We demonstrate the quantum advantage  of parameter saving using the amplitude encoding. The hybrid TN-VQC architecture provides a natural way to perform efficient compression of the input dimension, enabling further quantum RL applications on noisy intermediate-scale quantum devices.

\end{abstract}

\maketitle


\section{\label{sec:Indroduction}Introduction}
%

Reinforcement learning (RL) \cite{sutton2018reinforcement} has demonstrated remarkable abilities in achieving better than human performances on video games~\cite{Mnih2015Human-levelLearning, schrittwieser2019mastering, badia2020agent57, kapturowski2018recurrent}, and  the game of Go~\cite{silver2016mastering, silver2017mastering}.
Recently it has also been applied to quantum physics research such as quantum control~\cite{Fosel:2018wc, kuo2021quantum, sivak2021model}, quantum error correction~\cite{Sweke:2018hs,Liu:2018xi,Poulsen-Nautrup:2018jy}, quantum experiment design~\cite{Melnikov:2018vq}.
%
%
%
%
Meanwhile, quantum computers (QC), which theoretically promises significant speedup over classical computers on certain problems \cite{harrow2017quantum, nielsen2002quantum, shor1999polynomial, grover1997quantum}, have been realized recently \cite{grzesiak2020efficient, arute2019quantum, cross2018ibm}. However, these so-called noisy intermediate-scale quantum (NISQ) devices \cite{preskill2018quantum} lack quantum error correcting capabilities \cite{gottesman1997stabilizer,gottesman1998theory,nielsen2002quantum} and cannot perform quantum algorithms with a large number of qubits and a deep circuit. Such limitations make the development of near-term quantum algorithms highly non-trivial.

Numerous efforts have been made to utilize these NISQ resources and one of the notable achievements is the \emph{variational quantum algorithms} (VQA) \cite{cerezo2020variational, bharti2021noisy}. In such a framework,  certain parts of a given computational task that can leverage the strength of quantum physics will be put on the quantum computer, while the rest remains on the classical computer. The outputs from the quantum computer will be channeled into the classical computer and a predefined algorithm will determine how to adjust the parameters of the quantum circuit on the quantum computer. 

%
%
In classical RL,  evolutionary optimization has been shown to reach similar or even superior performance compared to gradient-based methods on certain difficult RL problems~\cite{such2017deep}. Therefore, it is natural to consider the potential application of this method in the quantum RL. To the best of our knowledge, applying evolutionary algorithms to quantum RL optimization has not been studied extensively. On the other hand, current quantum devices are realized with a small number of qubits, forbidding potential use cases of environments with large dimensions.
Here we present an evolutionary and gradient-free method to optimize the quantum circuit parameters for RL agents. We show that this method can successfully train the quantum deep RL model to achieve the state-of-the-art result on the Cart-Pole problem with only a few parameters, demonstrating a potential quantum advantage. In addition, we demonstrate that the evolutionary method can be used to optimize models combining tensor network (TN) and variational quantum circuits (VQC) in an end-to-end manner, opening up more opportunities for the application of quantum RL with NISQ devices.
In this work, we present an evolutionary deep quantum  RL framework to demonstrate the potential quantum advantage. 
Our contributions are:
\begin{itemize}
    \item Demonstrate the quantum advantage of parameter saving via amplitude encoding. In the Cart-Pole environment, we successfully use the amplitude encoding to encode a $4$-dimensional input vector into a two-qubit system.
    \item Demonstrate the capabilities of the hybrid TN-VQC architecture in quantum RL scenarios. The hybrid architecture can efficiently compress the large dimensional input into a small representation that could be processed with a NISQ device.
\end{itemize}

The paper is organized as follows. In Section~\ref{sec:ReinforcementLearning}, we introduce the basics of reinforcement learning. In Section~\ref{sec:TestingEnvironments}, we describe the testing environments used in this work. In Section~\ref{sec:VariationalQuantumCircuits}, we introduce  the basics of variational quantum circuits. Section~\ref{sec:QCForCartPole} describes the quantum circuit architecture for the Cart-Pole problem. Section~\ref{sec:HybridTNVQC} introduces tensor network methods and describes the hybrid TN-VQC architecture for the MiniGrid problem. Section~\ref{sec:QuantumCircuitEvolution} explains the evolutionary method used to optimize quantum circuit parameters. The performances of the proposed models are shown in Section~\ref{sec:ExpAndResults}, followed by further discussions in Section~\ref{sec:Discussion}. Finally we conclude our work in Section~\ref{sec:Conclusion}.

\section{\label{sec:ReinforcementLearning}Reinforcement Learning}
\emph{Reinforcement learning} is a machine learning paradigm where a given goal is to be achieved through an \emph{agent} interacting with an \emph{environment} $\mathcal{E}$ over a sequence of discrete time steps~\cite{sutton2018reinforcement}. At each time step $t$, the agent observes a \emph{state} $s_t$ and subsequently selects an \emph{action} $a_t$ from a set of possible actions $\mathcal{A}$ according to its current \emph{policy} $\pi$. The policy is a mapping from a certain state $s_t$ to the probabilities of selecting an action from $\mathcal{A}$. 
After performing the action $a_t$, the agent receives a scalar \emph{reward} $r_t$ and the state of the next time step $s_{t+1}$. For episodic tasks, the process proceeds over a number of time steps until the agent reaches the terminal state.
An \emph{episode} includes all the states the agent experienced throughout the aforementioned process, from a random selected initial state to the terminal state.
Along each state $s_t$ during the training process, the agent's overall goal is to maximize the expected return, which is quantified by the value function at state $s$ under policy $\pi$, $V^\pi(s) = \mathbb{E}\left[R_t|s_t = s\right]$, where $R_t = \sum_{t'=t}^{T} \gamma^{t'-t} r_{t'}$ is the \emph{return}, the total discounted reward from time step $t$. The discount factor $\gamma \in (0,1]$ allows the investigator to control the influence of future rewards on the agent's decision making. A large discount rate $\gamma$ forces the agent to take into account the farther future, whereas a small $\gamma$ allows the agent to focus more on immediate rewards and ignore future rewards beyond a few time steps. The value function can be expressed as $V^\pi(s) = \sum_{a\in\mathcal{A}} Q^\pi (s,a) \pi(a|s)$, where the \emph{action-value function} or \emph{Q-value function} $ Q^\pi (s,a) = \mathbb{E}[R_t|s_t = s, a]$ is the expected return of choosing an action $a \in \mathcal{A}$ in state $s$ according to the policy $\pi$. Selecting the best policy among all possible policies yields the maximal action-value, given by the optimal action-value function $Q^*(s,a) = \max_{\pi} Q^\pi(s,a)$, which in turn produces the maximal expected return.
\section{\label{sec:TestingEnvironments}Testing Environments}
\subsection{\label{sec:intro_cartpole}Cart-Pole}
\begin{figure}[htbp]
\center
\includegraphics[width=0.5\linewidth]{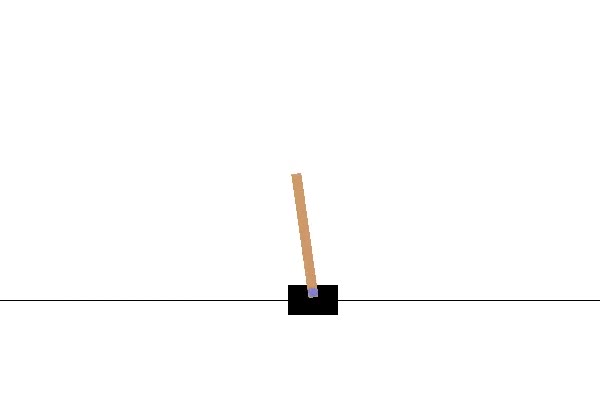}
\caption[Environment: Cart-Pole]{{\bfseries The Cart-Pole environment from OpenAI Gym.}
 }
\label{CartPole_Env}
\end{figure}

We first study the performance of a simple VQC model with the classic Cart-Pole problem, demonstrating the validity of quantum RL. Cart-Pole is a common testing environment for benchmarking simple RL models, and has been a standard example in the OpenAI Gym \cite{brockman2016openai} (see~\figureautorefname{\ref{CartPole_Env}}).
In this environment, a pole is attached by a fixed joint to a cart moving horizontally along a frictionless track. The pendulum initially stays upright, and the goal is to keep it as close to the initial state as possible by pushing the cart leftwards and rightwards.
The RL agent learns to output the appropriate action according to the observation it receives at each time step. \\
\textbf{The Cart-Pole environment mapping is:}
\begin{itemize}
\item Observation: A four dimensional vector $s_t$ comprising values of the {cart position, cart velocity, pole angle, and pole velocity at the tip}.
\item Action: There are two actions $+1$ and $-1$ in the action space, corresponding to pushing the cart rightwards and leftwards, respectively. How to choose the action with a variational quantum circuit is described in Sec.~\ref{sec:action_selection_cart_pole}. 
\item Reward: A reward of $+1$ is given for every time step where the pole remains close to being upright. An episode terminates if the pole is angled over $15$ degrees from vertical, or the cart moves away from the center more than $2.4$ units.
\end{itemize} 

\subsection{\label{sec:intro_minigrid} MiniGrid}
\begin{figure}[tbp]
\center
\includegraphics[width=0.8\linewidth]{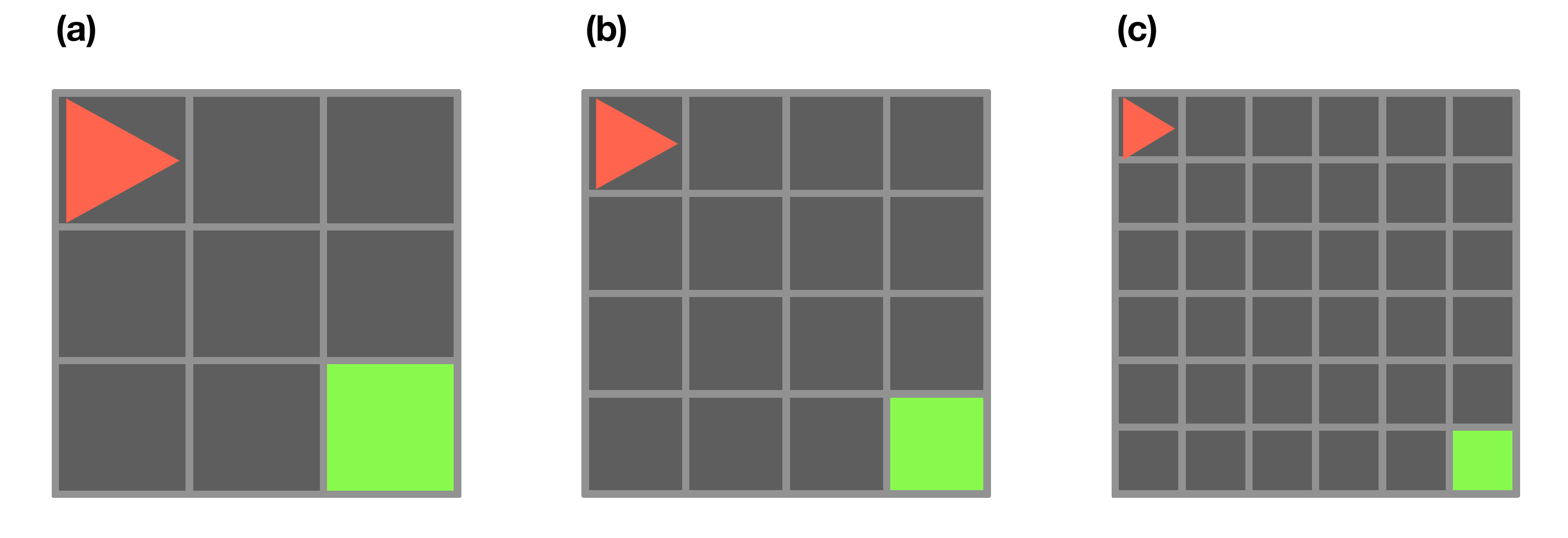}
\caption[Environment: MiniGrid]{{\bfseries The empty environment from MiniGrid.} The three environments from MiniGrid we consider in this work. In each environment, there are also walls which span $1$ unit on each side (not shown in the figure). (a), (b) and (c) represent the \texttt{MiniGrid-Empty-5x5-v0}, \texttt{MiniGrid-Empty-6x6-v0} and \texttt{MiniGrid-Empty-8x8-v0} environments, respectively.
 }
\label{MiniGridEmpty_Env}
\end{figure}

We study the model performance of our hybrid TN-VQC architecture (as described in Sec.~\ref{sec:HybridTNVQC}) with a more complex environment MiniGrid \cite{gym_minigrid}.
This environment also follows the standard OpenAI Gym API but has a much larger observation input, thus being a desirable choice for studying our hybrid RL model. In this environment, the RL agent receives a $7 \times 7 \times 3 = 147$ dimensional vector from observation and has to determine accordingly the action from the action space $\mathcal{A}$, which contains a total of six possibilities. As shown in \figureautorefname{\ref{MiniGridEmpty_Env}}, the agent (shown in red triangle) is expected to find the shortest path from the starting point to the goal (shown in green).\\
\textbf{The MiniGrid environment mapping is:}
\begin{itemize}
    \item Observation: A $147$ dimensional vector $s_t$.
    \item Action: There are six actions {$0$,$\cdots$,$5$} in $\mathcal{A}$, each corresponding to  the following,
    \begin{itemize}
        \item Turn left
        \item Turn right
        \item Move forward
        \item Pick up an object
        \item Drop the object being carried
        \item Toggle (open doors, interact with objects)
    \end{itemize}
    \item Reward: A reward of $1$ is given when the agent reaches the goal. A penalty is subtracted from the reward according to the formula:
    \begin{equation}
    1 - 0.9 \times (\textit{number of steps}/\textit{max steps allowed})    
    \end{equation}
    The \emph{max steps allowed} is defined to be $4 \times n \times n$ where $n$ is the grid size \cite{gym_minigrid}. In the present study, we consider $n = 5,6,8$. Such a reward scheme is challenging since it is \emph{sparse}, i.e. the agent will not receive any reward along the steps until it reaches the goal and therefore most of the actions elicit no immediate response from the environment.
\end{itemize}
\section{\label{sec:VariationalQuantumCircuits}Variational Quantum Circuits}
\begin{figure}[tbp]
\begin{center}
\begin{minipage}{10cm}
\Qcircuit @C=1em @R=1em {
\lstick{\ket{0}} & \multigate{3}{U(\mathbf{x})}  & \qw        & \multigate{3}{V(\boldsymbol{\theta})}       & \qw      & \meter \qw \\
\lstick{\ket{0}} & \ghost{U(\mathbf{x})}         & \qw        & \ghost{V(\boldsymbol{\theta})}              & \qw      & \meter \qw \\
\lstick{\ket{0}} & \ghost{U(\mathbf{x})}         & \qw        & \ghost{V(\boldsymbol{\theta})}              & \qw      & \meter \qw \\
\lstick{\ket{0}} & \ghost{U(\mathbf{x})}         & \qw        & \ghost{V(\boldsymbol{\theta})}              & \qw      & \meter \qw \\
}
\end{minipage}
\end{center}
\caption{{\bfseries Generic architecture for variational quantum circuits (VQC).}
$U(\mathbf{x})$ is the quantum circuit block for encoding the (classical) input data $\mathbf{x}$ into a quantum state and $V(\boldsymbol{\theta})$ is the variational block with  learnable parameters $\boldsymbol{\theta}$ to be optimized via gradient-based or gradient-free methods.
Quantum measurements are performed over some or all of the qubits.
}
\label{Fig:GeneralVQC}
\end{figure}
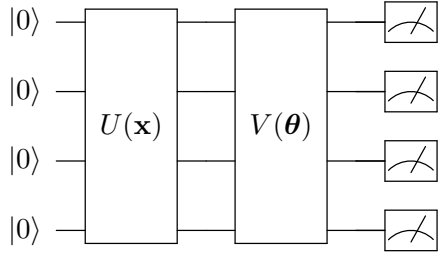

%
%
Variational quantum circuits have been established in quantum computing as a type of quantum circuits with parameters tunable via iterative optimizations, which are typically implemented with either gradient-based \cite{schuld2019evaluating, kyriienko2021generalized} or gradient-free methods \cite{franken2020gradient}. In the present study, we employ the gradient-free approach based on the evolutionary algorithm. The architecture of a generic VQC is illustrated in \figureautorefname{\ref{Fig:GeneralVQC}}. The $U(\mathbf{x})$ block serves as the state preparation part, which encodes the classical data $\mathbf{x}$ into the circuit quantum states and is not subject to optimization, whereas the $V(\boldsymbol{\theta})$ block represents the variational part containing \emph{trainable} parameters $\boldsymbol{\theta}$ , which in this study are optimized through evolutionary methods.
The output information is obtained through measuring a subset or all of the qubits and thereby retrieving a classical bit string.

One of the notable advantages of such circuits is that they are robust against quantum noise \cite{kandala2017hardware,farhi2014quantum,mcclean2016theory}, thus being potentially favorable for the NISQ devices.
In the field of quantum machine learning, applications of VQCs to standard machine learning tasks have achieved various degrees of success. Prominent examples include function approximation \cite{chen2020quantum, mitarai2018quantum,paine2021quantum,kyriienko2021solving}, classification \cite{mitarai2018quantum,schuld2018circuit,havlivcek2019supervised,Farhi2018ClassificationProcessors,benedetti2019parameterized,mari2019transfer, abohashima2020classification, easom2020towards, sarma2019quantum, stein2020hybrid,chen2020hybrid,chen2020qcnn,wu2020application,stein2021quclassi,chen2021hybrid,jaderberg2021quantum,mattern2021variational,hur2021quantum}, generative modeling \cite{dallaire2018quantum,stein2020qugan, zoufal2019quantum, situ2018quantum,nakaji2020quantum}, deep reinforcement learning \cite{chen19,lockwood2020reinforcement,jerbi2019quantum,Chih-ChiehCHEN2020,wu2020quantum,skolik2021quantum,jerbi2021variational,kwak2021introduction}, sequence modeling \cite{chen2020quantum, bausch2020recurrent, takaki2020learning}, speech recognition \cite{yang2020decentralizing}, metric and embedding learning \cite{lloyd2020quantum, nghiem2020unified}, transfer learning \cite{mari2019transfer} and federated learning \cite{chen2021federated,yang2020decentralizing,chehimi2021quantum}.
Furthermore, it has been shown that the VQCs may have more expressive power than classical neural networks \cite{sim2019expressibility,lanting2014entanglement,du2018expressive}. The \emph{expressive power} here is defined by the ability to represent certain functions or distributions given a limited number of parameters. 
Indeed, artificial neural networks (ANN) are known as \emph{universal approximators} \cite{hornik1989multilayer}, i.e. a neural network with even only one single hidden layer can in principle approximate any computable function. However, as the complexity of the function grows, the number of neurons required in the hidden layer(s) may become extremely large, posing considerable demands of computational resources. Whether VQCs can perform better than their classical counterparts with an equal or fewer number of parameters is therefore an interesting  subject to explore.

\subsection{Quantum Encoding}
For a quantum circuit to process data,
the classical input vector has to be encoded into a quantum state first.
A general $N$-qubit quantum state can be written as:
\begin{equation}
\label{eqn:quantum_state_vec}
    \ket{\psi} = \sum_{(q_1,q_2,...,q_N) \in \{ 0,1\}^N}^{} c_{q_1,q_2,...,q_N}\ket{q_1} \otimes \ket{q_2} \otimes ... \otimes \ket{q_N},
\end{equation}
where $ c_{q_1,q_2,...,q_N}$ are the \emph{amplitudes} of the quantum states with $ c_{q_1,q_2,...,q_N} \in \mathbb{C}$ and $q_i \in \{0,1\}$. 
The absolute square of each amplitude $c_{q_1,q_2,...,q_N}$ represents the \emph{probability} of measuring the state  $\ket{q_1} \otimes \ket{q_2} \otimes ... \otimes \ket{q_N}$, and all the probabilities sum up to $1$, i.e.
\begin{equation} 
\label{eqn:quantum_state_vec_normalization_condition}
\sum_{(q_1,q_2,...,q_N) \in \{ 0,1\}^N}^{} ||c_{q_1,q_2,...,q_N}||^2 = 1. 
\end{equation}
There are several kinds of encoding schemes commonly used in quantum ML applications \cite{Schuld2018InformationEncoding}. Different encoding methods provide varying extent of quantum advantage. Some of them are not readily implemented on real quantum hardwares due to the large circuit depth. 
In this work, we employ two different encoding schemes based on the problems of interest. We use \emph{amplitude encoding} (described in \sectionautorefname{\ref{sec:AmplitudeEncoding}}) for the Cart-Pole problem and \emph{variational encoding} for (described in \sectionautorefname{\ref{sec:VariationalEncoding}}) the MiniGrid problem.

\section{\label{sec:QCForCartPole}Quantum Architecture for the Cart-Pole Problem}
\begin{figure}[htbp]
\begin{center}
\begin{minipage}{10cm}
\Qcircuit @C=1em @R=1em {
\lstick{\ket{0}} & \multigate{1}{U(\mathbf{x})}  & \qw        & \ctrl{1}       & \gate{R(\alpha_1, \beta_1, \gamma_1)} & \qw      & \meter \qw \\
\lstick{\ket{0}} & \ghost{U(\mathbf{x})}         & \qw        & \targ          & \gate{R(\alpha_2, \beta_2, \gamma_2)} & \qw      & \meter \qw \gategroup{1}{4}{2}{5}{.7em}{--} 
}
\end{minipage}
\end{center}
\caption[Quantum circuit architecture for the Cart-Pole problem.]{{\bfseries Quantum circuit architecture for the Cart-Pole problem.}
 $U(\mathbf{x})$ is the quantum routine for amplitude encoding and the grouped box is the variational circuit block with tunable parameters $\alpha_i$, $\beta_i$, $\gamma_i$. In the evolutionary quantum RL architecture for the Cart-Pole problem, the grouped box repeats $4$ times. The total number of parameters is $2 \times 3 \times 4 + 2= 26$ where the extra $2$ parameters are the bias added after the measurement.
}
\label{Fig:VQCForCartPole}
\end{figure}
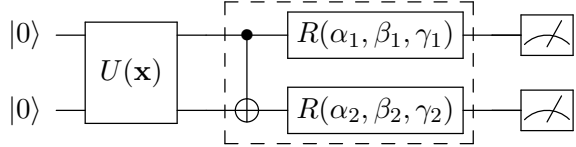

In this problem, the observation input is four-dimensional, which can be readily encoded into a quantum circuit. 
The quantum circuit for the Cart-Pole experiment is shown in \figureautorefname{\ref{Fig:VQCForCartPole}}.
In this $2$-qubit architecture, we load the $4$-dimensional observation input with \emph{amplitude encoding}. 
$U(\bm{x})$ represents the quantum routine for amplitude encoding, of which possible implementations is described in~\cite{Schuld2018InformationEncoding,mottonen2005transformation}. Here we repeat the variational circuit block (shown in grouped box) $4$ times to increase the number of parameters and thereby the expressive power.
In the measurement part, since there are only $2$ possible actions in the Cart-Pole problem, we simply evaluate the $Z$ expectation values of the two qubits individually. The measurement output, i.e. the expectation values, are then further processed by adding a classical \emph{bias} of the same dimension ($2$). 

\subsection{\label{sec:AmplitudeEncoding}Amplitude Encoding}
As the observation space of the Cart-Pole environment is continuous, it is impossible to use the computational basis encoding (which is for discrete space, as used in the previous work \cite{chen19}) to encode the input state. In this task, we employ the \emph{amplitude encoding} method to transform the observation into the amplitudes of a quantum state.
\emph{Amplitude encoding} is a method to encode a vector $(\alpha_{0}, \cdots, \alpha_{2^N-1})$ into an $N$-qubit quantum state 
$\ket{\Psi} = \alpha_{0}\ket{00\cdots 0} + \cdots + \alpha_{2^N-1}\ket{11\cdots 1}$ where the $\alpha_i$ are real numbers and the vector $(\alpha_{0}, \cdots, \alpha_{2^N-1})$ is normalized. A potential advantage is that for a $m$-dimensional vector, it requires only $\log_{2}(m)$ qubits to encode the data. The details of this operation is described in \appendixautorefname{\ref{appen:amplitude_encoding}}. The whole quantum circuit simulation is performed with the package \texttt{PennyLane}. 

\subsection{\label{sec:action_selection_cart_pole}Action Selection}
The selection of next action is similar to that used in the quantum deep $Q$-learning in the work \cite{chen19}. Specifically, the output from the quantum circuit after the classical post-processing of this $2$-qubit system is a $2$-tuple $[a,b]$. If $a$ is the maximum between the two values, then the corresponding action is $-1$; on the other hand, if the maximum is $b$, then the action is $+1$.
%
%
%
\section{\label{sec:HybridTNVQC}Hybrid TN-VQC Architecture for the MiniGrid Problem}
\begin{figure}[tbp]
\center
\includegraphics[width=.8\linewidth]{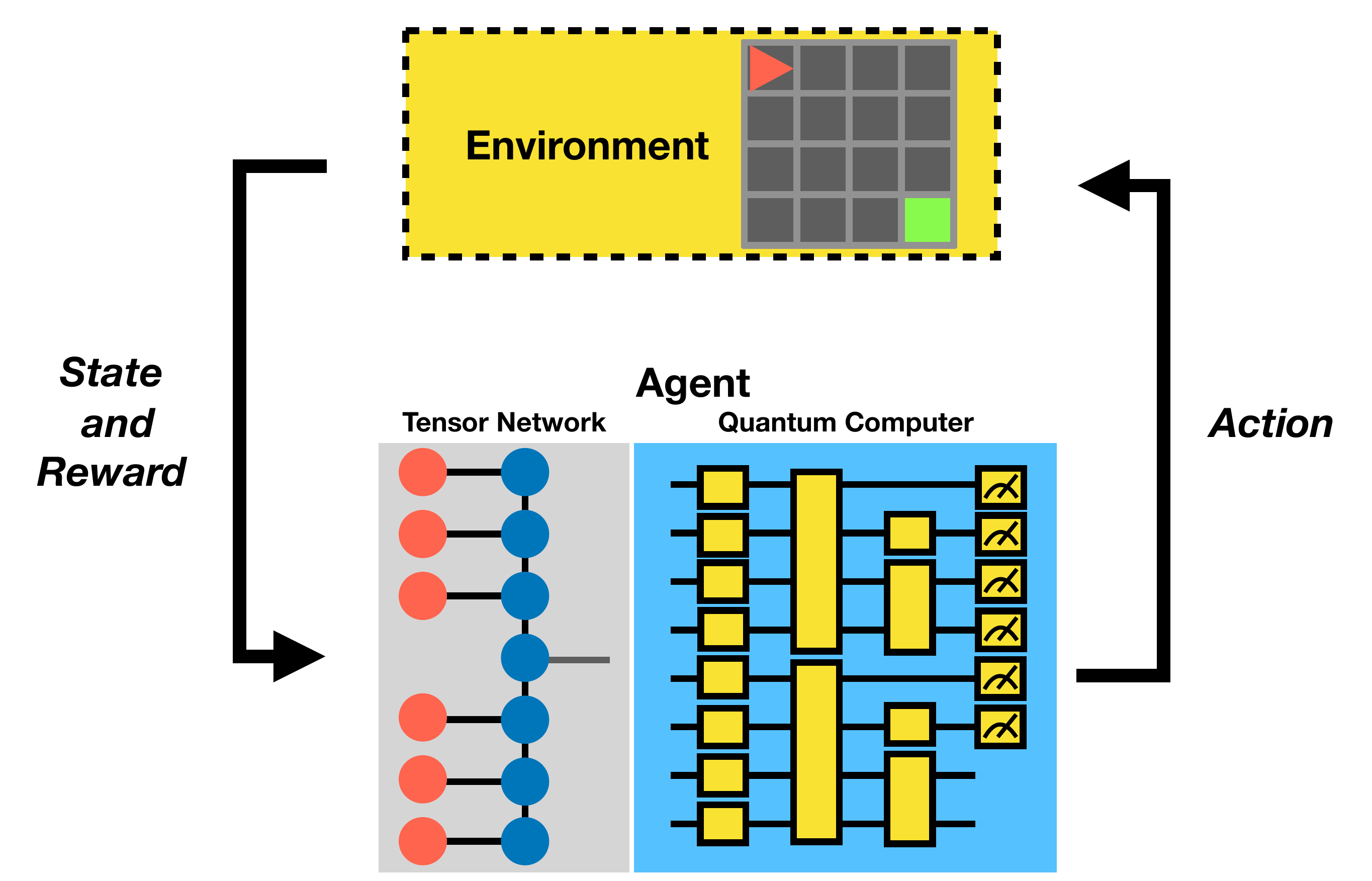}
\caption[Hybrid TN-VQC agent.]{{\bfseries The hybrid TN-VQC agent for the MiniGrid environment.} In this framework, the observation or state, which is of a dimension exceeding the capabilities of available quantum computers, is first compressed into a vector via the trainable MPS. The compressed representation is then encoded into the VQC for further training. The action is chosen based on the output of the VQC. In this work, we train such a TN-VQC model in an end-to-end fashion.
}
\label{HybridTN_VQC_RL_Diagram}
\end{figure}

One of the key challenges in the NISQ era is that  quantum computing machines are typically equipped with a limited number of qubits and can only execute quantum algorithms with a small circuit-depth. To process data with input dimension exceeding the number of available qubits, it is necessary to apply certain kinds of dimensional reduction techniques to first compress the input data. For example, in Ref.~\cite{mari2019transfer}, the authors applied a classical pre-trained convolution neural network to reduce the input dimension and then use a small VQC model to classify the images. However, the pre-trained model is already sufficiently powerful and it is not clear whether the VQC plays a critical role in the whole process. 
%
%
On the other hand, in Ref.~\cite{chen2020hybrid}, the authors explore the possibilities of using a TN for feature extraction and training the TN-VQC hybrid model in an end-to-end fashion. It has been shown that such a hybrid TN-VQC architecture succeed in the classification tasks. However, to our best knowledge, the potential of such an architecture has not yet been explored in other machine learning tasks. 
Since in the MiniGrid environment, the observation is a 147 dimensional vector, which is impossible to process on current NISQ devices, we propose a hybrid TN-VQC agent architecture (see \figureautorefname{\ref{HybridTN_VQC_RL_Diagram}}) so that an efficient dimensional reduction can be achieved.
\subsection{\label{sec:TensorNetwork}Tensor Network}
\begin{figure}[!htbp]
\center
\includegraphics[width=.8\linewidth]{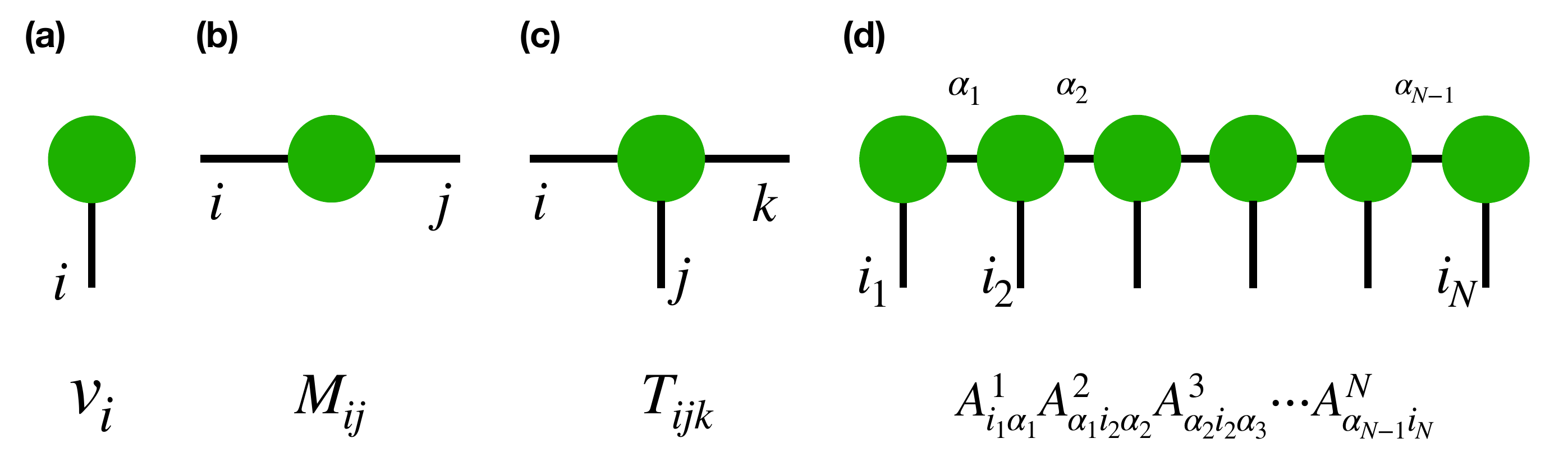}
\caption[Graphical Notation of Tensors and Tensor Networks]{{\bfseries Graphical Notation of Tensors and Tensor Networks.}
Graphical tensor notation of (a) a vector, (b) a matrix,  (c) a rank-3 tensor and (d) a MPS. Here we follow the Einstein convention that repeated indices, represented as closed legs in the diagram, are summed over.
 }
\label{TN_MPS_Diagram}
\end{figure}

\emph{Tensor networks} (TN) is a technique originally developed in the field of quantum many-body physics \cite{white1992density,white1993density,eisert2013entanglement,cirac2009renormalization,verstraete2008matrix,schollwock2005density,schollwock2011density} for efficiently expressing the quantum wave function $\ket{\Psi}$.
\emph{Matrix product states} (MPS), among others, is a type of one dimensional TN that decomposes a large tensor into a series of matrices. A general $N$-qubit quantum state can be written as
\begin{equation}
    \ket{\Psi} = \sum_{i_{1}}\sum_{i_{2}} \cdots \sum_{i_{N}} T_{i_{1}i_{2}\cdots i_{N}}\ket{i_{1}} \otimes \ket{i_{2}} \otimes \cdots \otimes \ket{i_{N}}
\end{equation}
where $T_{i_{1}i_{2}\cdots i_{N}}$ is the amplitude of each basis state $\ket{i_{1}} \otimes \ket{i_{2}} \otimes \cdots \otimes \ket{i_{N}}$. As the number of entries of $T_{i_{1}i_{2}\cdots i_{N}}$ grows exponentially with $N$, it is extremely difficult to store and process the quantum states on a classical computer when the number of qubits becomes large. However, with MPS, an efficient representation that approximately decomposes $T_{i_{1}i_{2}\cdots i_{N}}$ into a product of matrices \cite{MPS2007}:
\begin{equation}
\label{eqn:mps_definition}
    T_{i_{1}i_{2}\cdots i_{N}} = \sum_{\alpha_{1}}\sum_{\alpha_{2}} \cdots \sum_{\alpha_{N}} A_{i_{1} \alpha_{1}}^{1} A_{\alpha_{1} i_{2} \alpha_{2}}^{2} A_{\alpha_{2} i_{2} \alpha_{3}}^{3} \cdots A_{\alpha_{N-1} i_{N}}^{N},
\end{equation}
where the matrices $A$ are indexed from $1$ to $N$ and $\alpha_{j}$ represent the virtual indices, one can largely reduce the space where $\ket{\Psi}$ resides. Each \emph{virtual index} $\alpha_{j}$ has a dimension $m$ called \emph{bond dimension} and serves as a tunable hyperparameter in the MPS approximation. It is known that for a sufficiently large $m$, an MPS can represent any tensor \cite{verstraete2004density}. In machine learning applications, the bond dimension $m$ is typically used to 
tune the number of trainable parameters and thereby the expressive power of MPS.
\figureautorefname{\ref{TN_MPS_Diagram}} shows the illustration of tensors and MPS.
%
We refer to \cite{biamonte2017tensor} for in-depth introduction on tensor networks.
%

Since the pioneering work \cite{NIPS2016_6211}, great efforts have been made to apply TN in the field of machine learning. TN-based methods have been utilized for applications such as classification \cite{NIPS2016_6211, Liu_2019,Stoudenmire_2018,9058650,efthymiou2019tensornetwork, glasser2018supervised,bhatia2019matrix}, generative modeling \cite{PhysRevX.8.031012, PhysRevB.99.155131, PhysRevB.101.075135} and sequence modeling \cite{bradley2020modeling}. It has also been shown that TN-based architectures have deep connections to the building of quantum machine learning models \cite{huggins2019towards}. Specifically, it is possible to encode a quantum-inspired TN architecture such as MPS into a quantum circuit with single- and two-qubit gates \cite{ran2020encoding}.
\subsection{MPS Operation}
In the TN-VQC architecture in this study, we use a MPS-based feature extractor as the TN part to reduce the input dimension. For a MPS to process an input vector $\mathbf{v}$, a \emph{feature map} $\Phi(\mathbf{v})$ is needed. The general form is:
\begin{equation}
\mathbf{v} \rightarrow \ket{\Phi(\mathbf{v})} = \phi(v_{1}) \otimes \phi(v_{2}) \otimes \cdots \otimes \phi(v_{N}),
\end{equation}
where each $\phi$ is a $d$-dimensional feature map, mapping each $v_j$ into a $d$-dimensional vector. The value $d$ is known as the \emph{local dimension}. In this work, we choose $d = 2$ and the feature map $\phi(v_{j})$ to be:
\begin{equation}
    \phi(v_{j}) = \left[\begin{array}{c}1 - v_{j} \\ v_{j}\end{array}\right].
\end{equation}
The input vector $\mathbf{v}$, a state/observation perceived by the agent, is therefore encoded into a tensor product state in the following way,
\begin{equation}
\mathbf{v} \rightarrow \ket{\Phi(\mathbf{v})}=\left[\begin{array}{c}1 - v_{1} \\ v_{1}\end{array}\right] \otimes\left[\begin{array}{c}1 - v_2 \\ v_{2}\end{array}\right] \otimes \cdots \otimes\left[\begin{array}{c}1 - v_{N} \\ v_{N}\end{array}\right],
\end{equation}
%
%
which is then contracted with the trainable MPS and becomes a vector:
\begin{equation}
\label{eqn:loading_into_mps}
f(\mathbf{v})=\sum_{i_{1}}\sum_{i_{2}} \cdots \sum_{i_{N}} T_{i_{1} i_{2} \ldots i_{N}} \phi\left(v_{1}\right)_{i_{1}} \phi\left(v_{2}\right)_{i_{2}} \cdots \phi\left(v_{N}\right)_{i_{N}},
\end{equation}
where $i_{1}, i_{2}, \cdots, i_{N}$ are in $\{0,1\}$ and $T_{i_{1} i_{2} \ldots i_{N}}$ is defined as in \equationautorefname{\ref{eqn:mps_definition}} but with an additional rank-3 tensor in the middle with an open leg representing the 8-dimensional output, i.e. the compressed representation, as can be seen schematically in \figureautorefname{\ref{HybridTN_VQC_RL_Diagram}}. In \figureautorefname{\ref{HybridTN_VQC_RL_Diagram}}, the feature-mapped input and the trainable MPS are shown in red and blue circles (nodes), respectively. As the observation input in MiniGrid is a 147-dimensional vector, there are in total 147 input nodes and $(147+1)$ MPS nodes.
\subsection{\label{sec:VariationalEncoding}VQC Processing}
%
For the VQC part, we adopt the \emph{variational encoding} method to encode our compressed representations into quantum states.
The initial quantum state $\ket{0} \otimes \cdots \otimes \ket{0}$ first undergoes the $H \otimes \cdots \otimes H$ operation to become an unbiased state $\ket{+} \otimes \cdots \otimes \ket{+}$. Consider an $N$-qubit system, the corresponding unbiased state is,
\begin{equation}
\label{eqn:unbiasedInit}
\begin{split}
    \left( H\ket{0}\right)^{\otimes N} & =\underbrace{H\ket{0} \otimes \cdots \otimes H\ket{0}}_{N}\\
    & = \underbrace{\ket{+} \otimes \cdots \otimes \ket{+}}_{N}\\
    & = \underbrace{\left[\frac{1}{\sqrt{2}} \left(\ket{0} + \ket{1}\right)\right] \otimes \cdots \otimes \left[\frac{1}{\sqrt{2}} \left(\ket{0} + \ket{1}\right)\right]}_{N}\\
    & = \frac{1}{\sqrt{2^N}} \left(\ket{0} + \ket{1}\right)^{\otimes N} \\
    & = \sum_{(q_1,q_2,...,q_N) \in \{ 0,1\}^N} \frac{1}{\sqrt{2^N}} \ket{q_1} \otimes \ket{q_2} \otimes \cdots \otimes \ket{q_N}.
\end{split}
\end{equation}

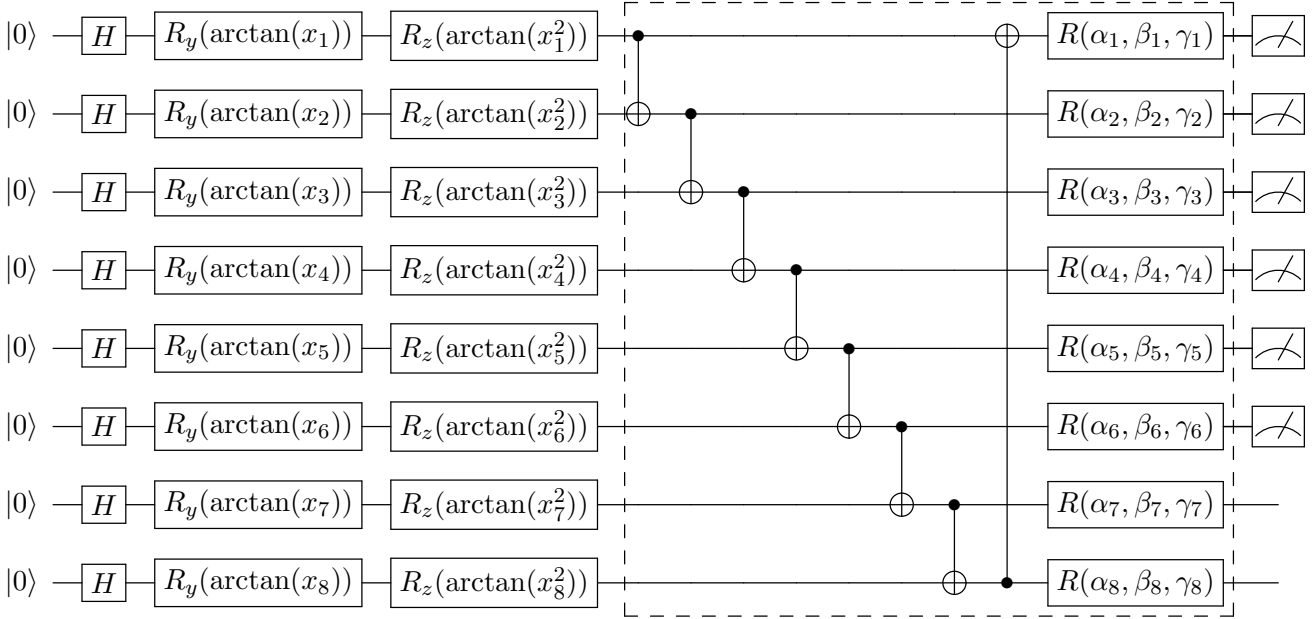
\begin{figure}
\begin{center}
\begin{minipage}{10cm}
\Qcircuit @C=1em @R=1em {
\lstick{\ket{0}} & \gate{H} & \gate{R_y(\arctan(x_1))} & \gate{R_z(\arctan(x_1^2))} & \ctrl{1}   & \qw       & \qw      & \qw      & \qw      & \qw      & \qw      & \targ    & \gate{R(\alpha_1, \beta_1, \gamma_1)} & \meter \qw \\
\lstick{\ket{0}} & \gate{H} & \gate{R_y(\arctan(x_2))} & \gate{R_z(\arctan(x_2^2))} & \targ      & \ctrl{1}  & \qw      & \qw      & \qw      & \qw      & \qw      & \qw      & \gate{R(\alpha_2, \beta_2, \gamma_2)} & \meter \qw \\
\lstick{\ket{0}} & \gate{H} & \gate{R_y(\arctan(x_3))} & \gate{R_z(\arctan(x_3^2))} & \qw        & \targ     & \ctrl{1} & \qw      & \qw      & \qw      & \qw      & \qw      & \gate{R(\alpha_3, \beta_3, \gamma_3)} & \meter \qw \\
\lstick{\ket{0}} & \gate{H} & \gate{R_y(\arctan(x_4))} & \gate{R_z(\arctan(x_4^2))} & \qw        & \qw       & \targ    & \ctrl{1} & \qw      & \qw      & \qw      & \qw      & \gate{R(\alpha_4, \beta_4, \gamma_4)} & \meter \qw \\
\lstick{\ket{0}} & \gate{H} & \gate{R_y(\arctan(x_5))} & \gate{R_z(\arctan(x_5^2))} & \qw        & \qw       & \qw      & \targ    & \ctrl{1} & \qw      & \qw      & \qw      & \gate{R(\alpha_5, \beta_5, \gamma_5)} & \meter \qw \\
\lstick{\ket{0}} & \gate{H} & \gate{R_y(\arctan(x_6))} & \gate{R_z(\arctan(x_6^2))} & \qw        & \qw       & \qw      & \qw      & \targ    & \ctrl{1} & \qw      & \qw      & \gate{R(\alpha_6, \beta_6, \gamma_6)} & \meter \qw \\
\lstick{\ket{0}} & \gate{H} & \gate{R_y(\arctan(x_7))} & \gate{R_z(\arctan(x_7^2))} & \qw        & \qw       & \qw      & \qw      & \qw      & \targ    & \ctrl{1} & \qw      & \gate{R(\alpha_7, \beta_7, \gamma_7)} &  \qw \\
\lstick{\ket{0}} & \gate{H} & \gate{R_y(\arctan(x_8))} & \gate{R_z(\arctan(x_8^2))} & \qw        & \qw       & \qw      & \qw      & \qw      & \qw      & \targ    & \ctrl{-7} & \gate{R(\alpha_8, \beta_8, \gamma_8)} & \qw \gategroup{1}{5}{8}{13}{.7em}{--}\qw 
}
\end{minipage}
\end{center}
\caption[Variational quantum circuit architecture for the action selection.]{{\bfseries Variational quantum circuit architecture for the action selection.}
The VQC component in our hybrid TN-VQC architecture includes three parts. The first one is the \emph{encoding} part, comprising Hadamard gates $H$ and single qubit rotation gates $R_y(\arctan(x_i))$ and $R_z(\arctan(x_i^2))$. The Hadamard gate $H$ serves to generate unbiased initial state as described in \equationautorefname{\ref{eqn:unbiasedInit}}, and the two rotation gates are for state preparation, where the input parameters $x_i$ are the compressed representation generated by the MPS component. The second one is the \emph{variational} part, comprising CNOT gates which entangle the quantum states of the qubits, and general rotation gates $R(\alpha,\beta,\gamma)$, where the parameters $\alpha_i$, $\beta_i$ and $\gamma_i$ are to be optimized via the learning process. The final one is the \emph{measurement} part, which gives the Pauli-$Z$ expectation values through multiple runs of the quantum circuit, where the quantum measurements are performed on the first $k$ qubits, $k$ being the number of actions. These retrieved values (logits) subsequently undergo classical processing such as the softmax function to yield the \emph{probability} of each possible action.}
\label{Fig:VQC_Hadamard_For_TN_QC_Hybrid}
\end{figure}
%
This unbiased quantum state subsequently goes through the \emph{encoding} part, which consists of $R_y$ and $R_z$ rotations. These rotation operations are parameterized by the compressed representation vector $\mathbf{x} = \left( x_1, x_2, \cdots, x_8 \right)$. For the $i^\text{th}$ qubit we choose the $R_y$ and $R_z$ rotation angles to be $\arctan(x_i)$ and $\arctan(x_i^2)$, respectively, where $i = 1, 2, \cdots, 8$. The encoded state is then processed with the \emph{variational} part with optimizable parameters, as shown in the dashed-box in \figureautorefname{\ref{Fig:VQC_Hadamard_For_TN_QC_Hybrid}}. The box can be repeated multiple times for increasing the number of parameters and thus the expressive power of the model. For the MiniGrid problems, we use only one block as a block number over one does not increase the performance significantly. In the final part of this model, the Pauli-$Z$ expectation values are retrieved through multiple runs (shots) of the circuit. Only the first 6 qubits are measured. These values are then processed with the softmax function for  evaluating the probabilities of each action.
%
\subsection{\label{sec:action_selection_minigrid}Action Selection}
The selection of next action is similar to that used in the quantum deep $Q$-learning in the work \cite{chen19}. Specifically, the output from the quantum circuit after the classical post-processing of this $8$-qubit system is a $6$-tuple $[a,b,c,d,e,f]$. If $a/b/c/d/e/f$ is the maximum among the six values, then the corresponding action is $0/1/2/3/4/5$.

%
%

\section{\label{sec:QuantumCircuitEvolution}Quantum Circuit Evolution}
Here we elucidate our quantum circuit evolution algorithm inspired by the work~\cite{such2017deep}. The essential concept of such an approach is to first generate a population of agents with random parameters and then make them evolve through a number of generations with certain  mutation rate. In each generation, the fittest agents will be selected for producing the next generation. The details of each step are explained below. See \appendixautorefname{\ref{appen:quantum_circuit_evolution}} for the pseudocode of the whole quantum circuit evolution algorithm.
\subsection{Initialization}
We first initialize the population $\mathcal{P}$ of $N$ agents with each of them given randomly generated initial parameters $\bm{\theta}$, which are sampled from the normal distribution $\mathcal{N}(0,I)$ and multiplied by a factor of $0.01$. The multiplication factor $0.01$ serves to set the parameters around zero, thereby rendering the training process more stable.
\subsection{Running the Agents}
For each generation, all of the agents' fitness are evaluated as follows. Each agent plays the game for $R_1$ times and the average score, which represents the fitness, is calculated by $S_{i}^{avg} = \frac{1}{R_1} \sum_{r = 1}^{R_1} S_{i,r}$ where $S_{i,r}$ is the score of the $r^\text{th}$ trial obtained by the $i^\text{th}$ agent. The score here is simply the sum of rewards within an episode.
Having the fitness of all the agents, we then select the top $T$ agents according to their average scores (fitness) $S_{i}^{avg}$. The resulting group is called the \emph{parents} and used to generate the next generation.
\subsection{Mutation and the Next Generation}
The $N$ children of the next generation are generated via two separate procedures.
The first part is to generate a group of $N-1$ children, each of which is a single agent randomly selected from the parent group and slightly mutated. Specifically, the parameter vector $\bm{\theta}$ of this parent agent undergoes the following \emph{mutation} operation: $\bm{\theta} \leftarrow \bm{\theta} + \sigma\epsilon$, where $\sigma$ is the \emph{mutation power} and $\epsilon$ the Gaussian noise sampled from the normal distribution $\mathcal{N}(0,I)$. This is distinct from the commonly used gradient-based methods in that the optimization direction is randomly chosen, a feature that can potentially provide the advantages of circumventing the local optima and efficiently optimizing the parameters in an environment with sparse rewards~\cite{such2017deep}.
The second part is to find the \emph{elite}, the $N^\text{th}$ child which is not mutated. To make the selection process more robust against noises, we make each agent from the parent group play the game $R_2$ times and obtain the average scores $E_{j}^{avg}$. The top $1$ agent, i.e. the one with the highest average score $E_{j}^{avg}$, is selected as the elite child.

\section{\label{sec:ExpAndResults}Experiments and Results}
We first demonstrate the quantum advantage of VQC with amplitude encoding in the standard benchmark Cart-Pole environment. Then we show the capabilities of our hybrid TN-VQC architecture in processing larger dimensional input state in the MiniGrid environment. The procedure of evolutionary optimization is the same for both experiments.
\subsection{Cart-Pole}
\begin{figure}[tbp]
\center
\includegraphics[width=1.\linewidth]{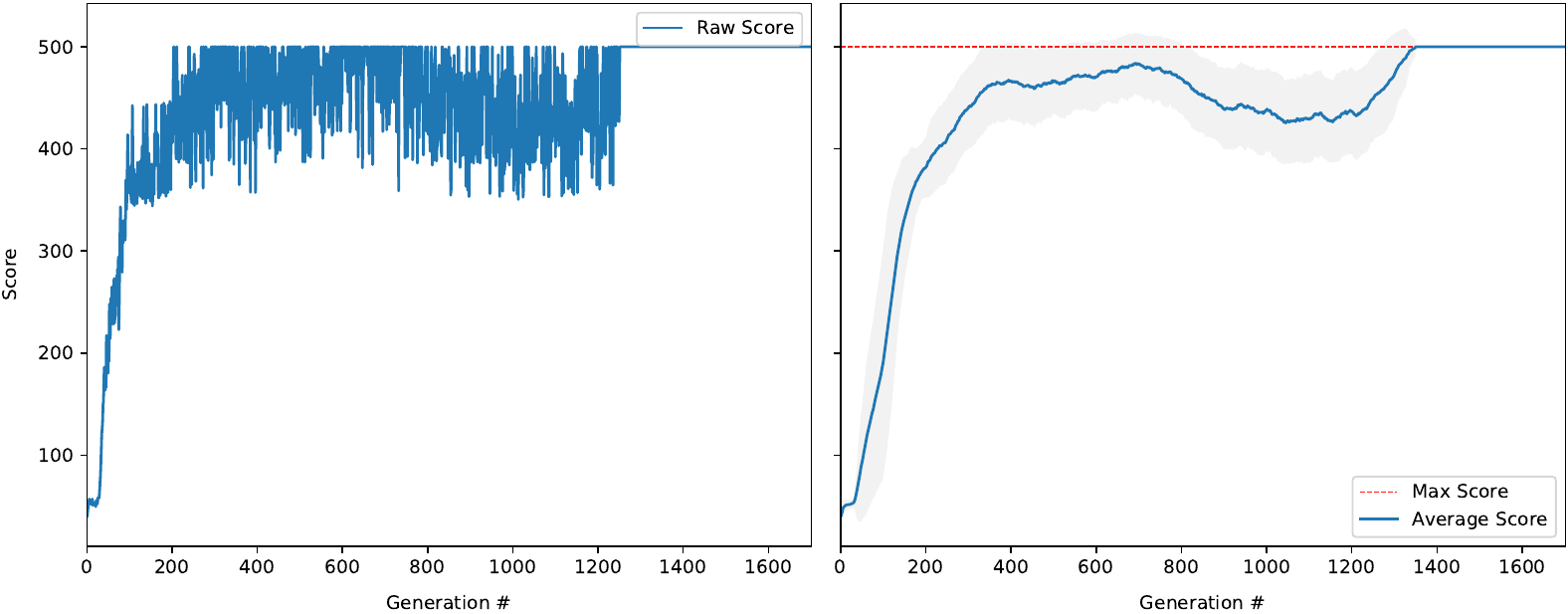}
\caption[Result:Evolutionary Quantum Deep Q-Learning on Cart-Pole]{{\bfseries Performance of the VQC  in the Cart-Pole environment.}
Left panel: The average score of the top-5 agents at each generation. The maximum steps allowed in this environment is $500$. Right panel: The gray area represents the standard deviation of rewards in each iteration during exploration with the standard RL reproducible setting for stability. The mean and the standard deviation values are calculated from the raw scores (rewards) in the past 100 generations.}
\label{VQDQN_Evo_CartPole_Result_Top}
\end{figure}

In this experiment, we set the number of generations to be $1700$, the population size $N = 500$, the truncation selection number $T = 5$, the mutation power $\sigma = 0.02$, the number of repetition (for evaluating all the agents) $R_1 = 3$ and the number of repetition (for evaluating the parents) $R_2 = 5$. 
The simulation results of this experiment are shown in~\figureautorefname{\ref{VQDQN_Evo_CartPole_Result_Top}}. It can be seen that after about $250$ generations, the average score of the top $5$ agents are steadily above $400$, and after around $1300$ generations, the top $5$ agents all converge to the optimal policy and reach a score of $500$, which corresponds to the maximum steps allowed in the environment. 

A notable achievement is that we use only $26$ parameters to reach the optimal result, which is less than those required in typical classical neural networks by at least one order of magnitude. Empowered by amplitude encoding as well as the nature of VQC, we significantly reduce the number of parameters in this specific problem. It is thus highly desirable to explore the feasibility of applying such an encoding method to other quantum RL problems, which could bring about quantum advantage in a way of reducing the model complexity as quantified by the number of parameters to as small as $\text{poly}(\log{n})$, in contrast to the $\text{poly}(n)$ parameters typically required in standard neural networks where $n$ is the dimension of the input vector.
\subsection{MiniGrid}
\begin{figure}[tbp]
\center
\includegraphics[width=1.\linewidth]{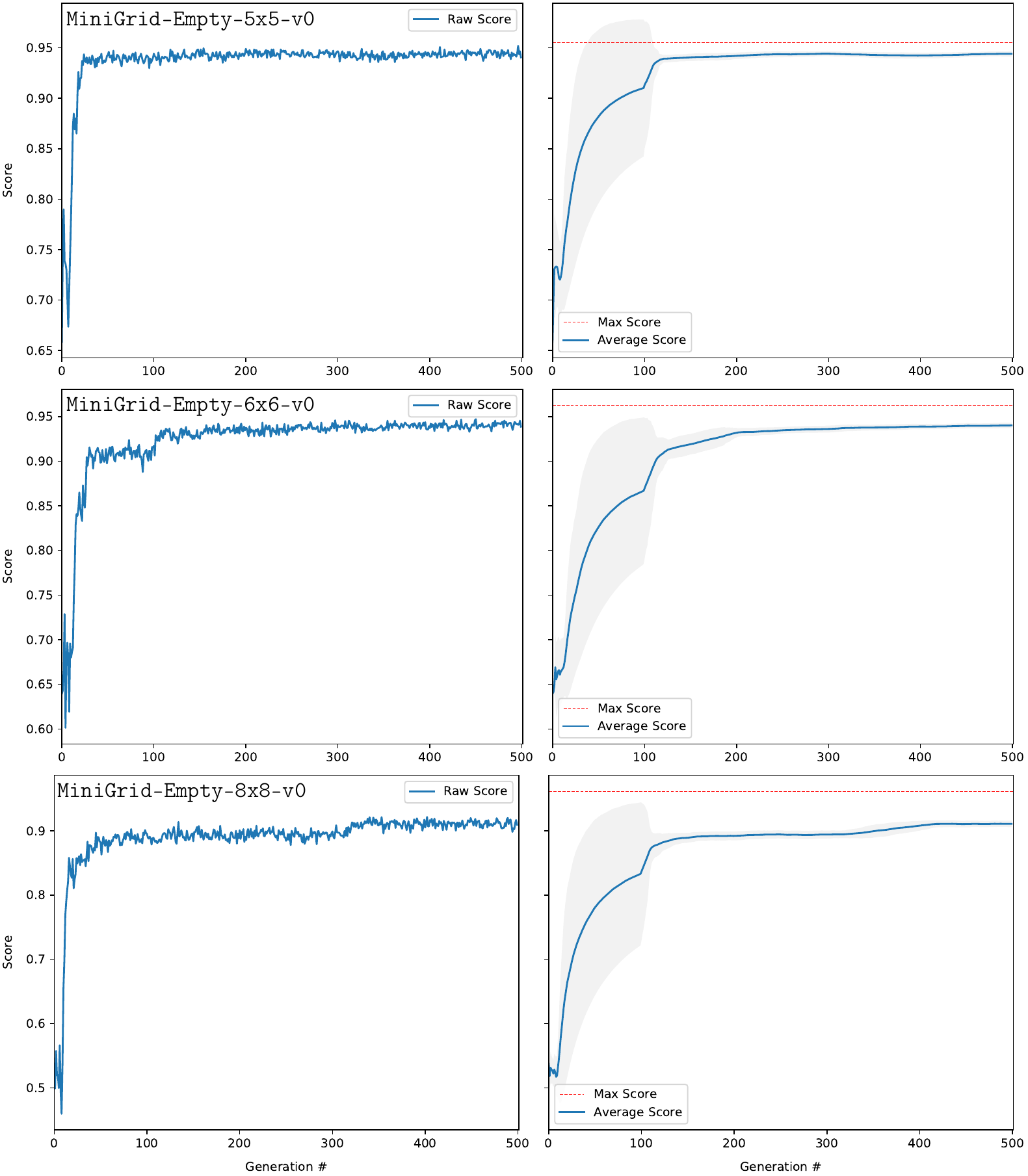}
\caption[Result:Evolutionary Quantum Deep Q-Learning on MiniGrid 5x5 Environment.]{{\bfseries Performance of the hybrid TN-VQC in the MiniGrid environments.}
Left panels: The average score of the top-5 agents at each generation. Right panels: The gray area represents the standard deviation of rewards in each iteration during exploration with the standard RL reproducible setting for stability. The mean and standard deviation values are calculated from the raw scores (rewards) in the past 100 generations.}
\label{VQDQN_Evo_MiniGrid_Result_Top}
\end{figure}
Here we consider  three configurations as shown in \figureautorefname{\ref{MiniGridEmpty_Env}}.
The observation is a $147$-dimensional vector and there are $6$ possible actions, as described in \sectionautorefname{\ref{sec:intro_minigrid}}.
%
In this experiment, we employ the hybrid TN-VQC architecture combining MPS and VQC. The MPS feature extractor receives the $147$-dimensional input state from the environment and output a $8$-dimensional vector to be encoded into the VQC. Empowered by the tensor network method, we successfully compress the large input vector into a small vector representation favorable for quantum circuit processing. This opens the possibilities of studying other complex RL problems with quantum circuits via applying such a dimension reduction technique.
%
We set the number of generations to be $500$, the population size $N = 500$, the truncation selection number $T = 10$, the mutation power $\sigma = 0.02$, the number of repetition (for evaluating all the agents) $R_1 = 3$ and the number of repetition (for evaluating the parents) $R_2 = 5$.
The performance of our hybrid TN-VQC model in the three MiniGrid environments \texttt{MiniGrid-Empty-5x5-v0}, \texttt{MiniGrid-Empty-6x6-v0} and \texttt{MiniGrid-Empty-8x8-v0} are shown in \figureautorefname{\ref{VQDQN_Evo_MiniGrid_Result_Top}}. 
%


In the \texttt{MiniGrid-Empty-5x5-v0} environment (with a maximum score of $0.955$), the simplest one of the three, it is clear that the average score of the top-5 agents is able to reach near-optimal value in less than 40 generations.
In the \texttt{MiniGrid-Empty-6x6-v0} environment (with a maximum score of $\sim0.956$), which is harder than the previous one, we observe that the average score of the top-5 agents rises above $0.9$ in 40 generations and reaches near-optimal value after around 120 generations.
In the \texttt{MiniGrid-Empty-8x8-v0} environment (with a maximum score of $\sim0.961$), the most difficult one among the three, we can see that it takes about 350 generations for the average score of the top-5 agents to rise and stay steadily above 0.9. It is clear that our model performs the worst in the last environment in terms of both the convergence speed and the final scores.

\section{\label{sec:Discussion}Discussion}

\subsection{Relevant Studies}
Early work on quantum reinforcement learning can be traced back to \cite{dong2008quantum}, which needs to load the environment into quantum superposition states. This is not generally applicable for \emph{classical} environments. In  \cite{dunjko2015framework}, the authors consider the situation where computational agents are coupled to environments which are quantum-mechanical. More recent studies introduce and facilitate the use of variational quantum circuits in the applications of reinforcement learning \cite{chen19, lockwood2020reinforcement, jerbi2021variational, skolik2021quantum, wu2020quantum}. In contrast to the present study, all of them use gradient-based methods to optimize the policy and/or value function. For the Cart-Pole testing environment which is studied in both works \cite{jerbi2021variational, skolik2021quantum}, we observed that these two works and ours all reach the optimal solution. However, our model only requires $2$ qubits, which is significantly smaller than the models in \cite{jerbi2021variational, skolik2021quantum}, which employ a $4$-qubit VQC architecture. An interesting research direction is to what extent the gradient-free or gradient-based actually affects the performance of VQC in RL problems. Could certain VQC architectures benefit more from a particular kind of optimization method? We leave this for future investigation. For a more detailed review of recent developments in quantum reinforcement learning, we refer the interested readers to \cite{Dunjko2017AdvancesLearning, jerbi2019quantum}.


\subsection{Complex Benchmarks}
In this work, we further extend the complexity of the testing environments in comparison to previous works, including the study \cite{chen19} which considers discrete observation/states. In particular, we largely push the boundary of quantum RL via incorporating the quantum-inspired tensor network into a VQC-based architecture. Despite the success, there is still a significant gap between the current quantum RL and the classical RL in terms of the capabilities of processing high-dimensional inputs.
\subsection{Evolutionary Quantum Circuits}
The use of evolutionary methods in optimizing quantum circuits can also be found in these recent works \cite{anand2020natural,franken2020gradient,lu2020markovian}. Both of the works \cite{franken2020gradient, anand2020natural} use evolutionary methods to optimize VQE problems. Notably, the work \cite{franken2020gradient} introduces an evolutionary approach involving structural mutation to optimize the quantum circuit.
On the other hand, the work \cite{lu2020markovian} utilizes graph-encoding method to encode a quantum circuit and then adopt an evolutionary method to optimize the quantum model for certain classification tasks.
However, none of these works considers the direct application of evolutionary optimization to quantum reinforcement learning problems. Our work not only demonstrates the first successful implementation of evolutionary method for quantum RL, but also touches on another aspect rarely studied: the end-to-end training of a hybrid model consisting of MPS and VQC.
\subsection{More from Classical Neuroevolution}
In the present study, we employ the evolutionary algorithms specifically for optimizing the quantum circuit parameters. In classical neuroevolution, the whole architectures as well as the neural network parameters can be optimized through evolution. A framework called \emph{NeuroEvolution of Augmenting Topologies} (NEAT) \cite{stanley2002efficient} has been proposed for evolving the classical neural network's topologies along with its weights. It is thus intriguing to investigate the prospect of applying such concepts to evolving quantum machine learning architectures. For evolving a model with a complex architecture and a sizable amount of parameters, it is crucial to encode the model itself in an efficient fashion. Recent advances in neuroevolution \cite{zhang2020evolving} could serve as a guidance for designing high-performance evolutionary algorithms in quantum ML. 

A major issue yet to be addressed in our work is that our hybrid model can only reach sub-optimal results on harder problems. In particular, it is challenging for the current model to achieve the maximum score when the rewards are sparse. One of the potential solutions to this issue is the \emph{novelty search} developed in classical neuroevolution \cite{lehman2010efficiently, risi2010evolving}. The idea behind novelty search is that the agent is not trained to achieve the \emph{objective} in a conventional way. Rather, novelty search rewards the agents that behave differently \cite{lehman2011abandoning}. In classical deep RL problems, novelty search has been shown capable of solving hard RL problems with sparse rewards \cite{conti2017improving}. A potential direction of future research is thus to investigate whether such framework works in the quantum regime.




Evolutionary algorithms also play a critical role in the security of RL models. For example, in the work \cite{yang2020enhanced}, the authors explore the potential of applying evolutionary algorithms to attacking a deep RL model. Hence, another potential research direction is to study the robustness of quantum RL agents. \cite{yang2021causal}

\subsection{Training on a Real Quantum Computer}
Given the fact that currently available quantum computers suffer seriously from noises, in this research we only consider the case of noise-free simulation. Although previous results \cite{kandala2017hardware, farhi2014quantum, mcclean2016theory} have indicated that VQC-based algorithms may be resilient to noises through the absorption of these undesirable effects into the tunable parameters, with the limitation of current cloud-based quantum computing resources, it is impractical to implement the whole training process on a real quantum computer to verify customized models such as the TN-VQC one we propose. We expect that such issues could be resolved when commercial quantum devices become more reliable and accessible.
\section{\label{sec:Conclusion}Conclusion}
In this study, we present two quantum reinforcement learning frameworks based on evolutionary algorithm, of which one is purely quantum and the other has a hybrid quantum-classical architecture. In particular, we study two input loading schemes that can reduce the required qubit number of the VQC: amplitude encoding and tensor network compression, and demonstrate through numerical simulation the performance of each.
First, through the Cart-Pole problem (with an input dimension of 4), we show that with amplitude encoding, a framework based on a VQC can provide quantum advantage in terms of parameter saving. With a proper design of the quantum circuit, it is possible to reduce the number of parameters to the scale of $\text{poly}(\log(n))$, where $n$ is the input dimension.
Second, through the more complex MiniGrid problems (with an input dimension of 147), with the incorporation of a TN-based dimensional reduction method, we  show the possibility of  compressing an input of larger dimension into a representation that can be readily encoded into currently available quantum devices. Notably, this hybrid TN-VQC model can be trained in an end-to-end fashion, i.e. the TN and VQC parts are trained as a whole.
The results in this work strongly suggest the prospect of such versatile and scalable frameworks, which could shed light on the future designing of reinforcement learning algorithms for near-term quantum computers.

%

\begin{acknowledgments}

This work is supported by the U.S.\ Department of Energy, Office of Science, Office of High Energy Physics program under Award Number DE-SC-0012704 and the Brookhaven National Laboratory LDRD \#20-024 (S.Y.-C.C.), the Ministry of Science and Technology (MOST) of Taiwan under grants No. 107-2112-M-002-016-MY3, 108-2112-M-002-020-MY3 (Y.-J.K.),  109-2112-M-002-023-MY3,   109-2627-M-002-003, 107-2627-E-002-001-MY3,  109-2622-8-002-003 (H.S.G.) and the U.S. Air Force Office of Scientific Research under award number FA2386-20-1-4033, and by the National Taiwan University under Grant No. NTU-CC-110L890102 (H.S.G.).
\end{acknowledgments}
\newpage
\appendix
\section{\label{appen:amplitude_encoding}Amplitude Encoding}
\begin{figure}[htbp]
\begin{center}
\begin{minipage}{10cm}
\Qcircuit @C=1em @R=1em {
\lstick{\ket{q_1}}       & \ctrlo{2}               & \qw                          & \ctrl{2}                     & \ctrlo{2}                   & \qw                         & \ctrl{2}                    & \gate{R_y(\beta_{1}^n)}     & \qw      & \qw \\
\raisebox{.3em}{\vdots}  &                         & \raisebox{.3em}{\ldots}      &                              &                             & \raisebox{.3em}{\ldots}     &                              &                             &          &      \\
\lstick{\ket{q_{n-1}}}   & \ctrlo{1}               & \qw                          & \ctrl{1}                     & \gate{Ry(\beta_{1}^2)}      & \qw                         & \gate{Ry(\beta_{2^{n-2}}^2)} & \qw                         & \qw      & \qw  \\
\lstick{\ket{q_n}}       & \gate{Ry(\beta_{1}^1)}  & \qw                          & \gate{Ry(\beta_{2^{n-1}}^1)} & \qw                         & \qw                         & \qw                          & \qw                         & \qw      & \qw
}
\end{minipage}
\end{center}
\caption[Generic Amplitude Encoding Circuit.]{{\bfseries Generic quantum routine for amplitude encoding (before inversion).}
}
\label{Fig:GenericAmpEncoding}
\end{figure}
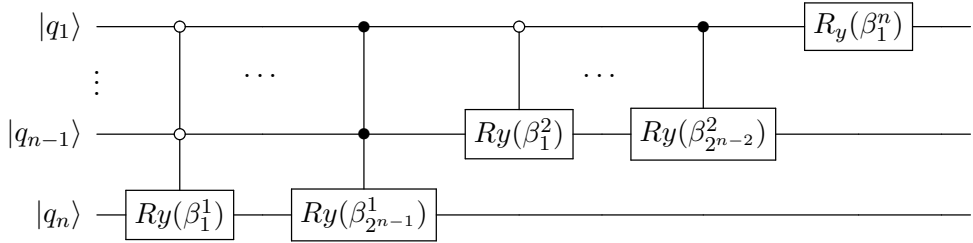

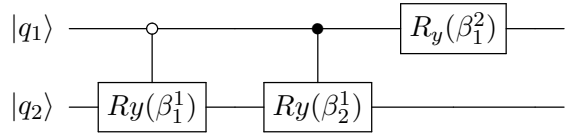
\begin{figure}[htbp]
\begin{center}
\begin{minipage}{10cm}
\Qcircuit @C=1em @R=1em {
\lstick{\ket{q_1}} & \ctrlo{1}  & \qw        & \ctrl{1}       & \gate{R_y(\beta_{1}^2)} & \qw      & \qw \\
\lstick{\ket{q_2}} & \gate{Ry(\beta_{1}^1)}  & \qw        & \gate{Ry(\beta_{2}^1)}      & \qw        & \qw      &  \qw
}
\end{minipage}
\end{center}
\caption[Amplitude Encoding Circuit for Cart-Pole.]{{\bfseries Quantum routine for amplitude encoding (before inversion) for the Cart-Pole experiment.}
}
\label{Fig:AmpForCartPole}
\end{figure}

Here we elucidate the quantum operation behind the \emph{amplitude encoding} used in the Cart-Pole experiment. We adopt the method presented in the work \cite{mottonen2005transformation}, which is further demonstrated for QML applications in the study \cite{Schuld2018InformationEncoding}.
The general goal of amplitude encoding  is to encode a vector $(\alpha_{0}, \cdots, \alpha_{2^n-1})$ into a $n$-qubit quantum state 
$\ket{\Psi} = \alpha_{0}\ket{00\cdots 0} + \cdots + \alpha_{2^n-1}\ket{11\cdots 1}$, where $\alpha_i$ are real numbers and the vector $(\alpha_{0}, \cdots, \alpha_{2^n-1})$ is normalized. This can be achieved by inversely running the quantum routine which transforms an arbitrary quantum state $\ket{\Psi} = \alpha_{0}\ket{00\cdots 0} + \cdots + \alpha_{2^n-1}\ket{11\cdots 1}$ into the state $\ket{00 \cdots 0}$.
In the work \cite{mottonen2005transformation}, the authors demonstrate the quantum routine to perform this transformation via a sequence of multi-controlled rotations. Here we follow the presentation in \cite{Schuld2018InformationEncoding}.
Consider a $n$-qubit system, the generic multi-controlled rotations around the vectors $v^i$ with angles $\beta_i$ on the last qubit $q_n$ consists of the sequential operations of the following $2^{n-1}$ gates:
\begin{equation}
\label{eqn:multiControlledRotation}
\begin{array}{l}
c_{q_{1}=0} \cdots c_{q_{n-1}=0} R_{q_{n}}\left(v^{1}, \beta_{1}\right) \quad\left|q_{1} \ldots q_{n-1}\right\rangle\left|q_{n}\right\rangle, \\ 
c_{q_{1}=0} \cdots c_{q_{n-1}=1} R_{q_{n}}\left(v^{2}, \beta_{2}\right) \quad\left|q_{1} \ldots q_{n-1}\right\rangle\left|q_{n}\right\rangle, \\
\mathrel{\makebox[\widthof{==}]{\vdots}} \\
c_{q_{1}=1} \cdots c_{q_{n-1}=1} R_{q_{n}}\left(v^{2^{n-1}}, \beta_{2^{n-1}}\right) \quad\left|q_{1} \ldots q_{n-1}\right\rangle\left|q_{n}\right\rangle.
\end{array}
\end{equation}
In the ML applications of interest, we consider only the case where  all the amplitudes are real numbers. In this scenario, the quantum routine includes a \emph{cascade} of multi-controlled $R_y$ rotations \cite{Schuld2018InformationEncoding}.
The \emph{cascade} of operations refers to sequentially implemented multi-controlled rotations on each qubit $q_i$ in the system. For example, the operation on the last qubit $q_n$ is shown in \equationautorefname{\ref{eqn:multiControlledRotation}}. 
In a $n$-qubit system (see~\figureautorefname{\ref{Fig:GenericAmpEncoding}} for illustration), the first operation in the cascade is the one on qubit $q_n$,
\begin{equation}
\label{eqn:multiControlledRotationLastQubit}
\begin{array}{l}
c_{q_{1}=0} \cdots c_{q_{n-1}=0} R_{q_{n}}\left(v^{1}, \beta_{1}^1\right) \quad\left|q_{1} \ldots q_{n-1}\right\rangle\left|q_{n}\right\rangle, \\ 
c_{q_{1}=0} \cdots c_{q_{n-1}=1} R_{q_{n}}\left(v^{2}, \beta_{2}^1\right) \quad\left|q_{1} \ldots q_{n-1}\right\rangle\left|q_{n}\right\rangle, \\
\mathrel{\makebox[\widthof{==}]{\vdots}} \\
c_{q_{1}=1} \cdots c_{q_{n-1}=1} R_{q_{n}}\left(v^{2^{n-1}}, \beta_{2^{n-1}}^1\right) \quad\left|q_{1} \ldots q_{n-1}\right\rangle\left|q_{n}\right\rangle.
\end{array}
\end{equation}
where the rotation axes $v^{1}, \cdots, v^{2^{n-1}}$ are $y$-axis ($R_y$ rotation).
The second operation in the cascade is implemented on the qubit $q_{n-1}$ in a similar way,
\begin{equation}
\label{eqn:multiControlledRotationSecondLastQubit}
\begin{array}{l}
c_{q_{1}=0} \cdots c_{q_{n-2}=0} R_{q_{n-1}}\left(v^{1}, \beta_{1}^2\right) \quad\left|q_{1} \ldots q_{n-2}\right\rangle\left|q_{n-1}\right\rangle, \\ 
c_{q_{1}=0} \cdots c_{q_{n-2}=1} R_{q_{n-1}}\left(v^{2}, \beta_{2}^2\right) \quad\left|q_{1} \ldots q_{n-2}\right\rangle\left|q_{n-1}\right\rangle, \\
\mathrel{\makebox[\widthof{==}]{\vdots}} \\
c_{q_{1}=1} \cdots c_{q_{n-2}=1} R_{q_{n-1}}\left(v^{2^{n-2}}, \beta_{2^{n-2}}^2\right) \quad\left|q_{1} \ldots q_{n-2}\right\rangle\left|q_{n-1}\right\rangle.
\end{array}
\end{equation}
The rotation angles $\beta_j^s$ can be shown to be \cite{mottonen2005transformation}
\begin{equation}
\beta_{j}^{s}=2 \arcsin \left(\frac{\sqrt{\sum_{l=1}^{2^{s-1}}\left|\alpha_{(2 j-1) 2^{s-1}+l}\right|^{2}}}{\sqrt{\sum_{l=1}^{2^{s}}\left|\alpha_{(j-1) 2^{s}+l}\right|^{2}}}\right),
\end{equation}
where $\alpha_i$ represents each of the amplitudes $(\alpha_{0}, \cdots, \alpha_{2^n-1})$.
To utilize the aforementioned quantum routine to perform \emph{amplitude encoding}, we can simply invert each and every operation and apply them in reverse order on the initial quantum state $\ket{00 \cdots 0}$ ~\cite{Schuld2018InformationEncoding}. We provide the example for $2$-qubit system used in our Cart-Pole experiment in~\figureautorefname{\ref{Fig:AmpForCartPole}}.

\section{\label{appen:quantum_circuit_evolution}Quantum Circuit Evolution Algorithm}
\begin{center}
\scalebox{0.75}{
\begin{minipage}{\linewidth}

\begin{algorithm}[H]
\begin{algorithmic}
\State Define the mutation power $\sigma$
\State Define the number of generation $M$
\State Define the number of truncation selection $T$
\State Define the repetition number of playing (all agents) $R_1$
\State Define the repetition number of playing (top agents)$R_2$
\State Initialize the population $\mathcal{P}$ with $N$ individuals/agents
\State Initialize action-value function quantum circuit $Q$ with random parameters $\bm{\theta}$ for each individual/agent
\For{generation $g=1,2,\ldots,M$} 
    \For {individual/agent $i=1,2,\ldots,N$}
        \For {$r = 1,2,\ldots,R_1$}
            \State Reset the testing environment
            \State Reset the cumulative reward (score) $S_{i,r} \leftarrow 0$
            \For {$t = 1,2,\ldots,K$}
            	\State The agent selects $a_t = \max_{a} Q^*(s_t, a; \bm{\theta})$ from the output of \\\hspace{2.5cm}
            	the quantum circuit
            	\State Execute action $a_t$ in emulator and observe reward $r_t$ and \\\hspace{2.5cm}
            	next state $s_{t+1}$
            	\State Record the reward $S_{i,r} \leftarrow S_{i,r} + r_t$
            \EndFor
        \EndFor
        \State Output the average score of the agent after playing $R_1$ times as \\\hspace{1.3cm}
        $S_{i}^{avg} = \frac{1}{R_1} \sum_{r = 1}^{R_1} S_{i,r}$
    \EndFor
    \State Output the top $T$ agents according to their average scores $S_{i}^{avg}$

    \For {$c = 1,2,\ldots,N-1$}
        \State Randomly select an agent from the top $T$ agents
        \State Mutate the parameter vector $\bm{\theta}$ according to the rule $\bm{\theta} \leftarrow \bm{\theta} + \sigma\epsilon$ 
        \\\hspace{1.3cm}
        where the mutation is Gaussian $\epsilon \sim \mathcal{N}(0,I)$
    \EndFor
    
    \For {agent in top $T$ agents $j = 1,2,\ldots,T$}
        \State Playing the game $R_2$ times
        \State Record the average score over playing $R_2$ times $E_{j}^{avg}$
    \EndFor
    \State Keep the best agent according to the average scores $E_{j}^{avg}$ as the $N$-th children\\\hspace{.6cm} which is the elite
    
\EndFor
\end{algorithmic}
\caption{Evolutionary Quantum Deep Q Learning}
\label{alg_evoRL}
\end{algorithm}
\end{minipage}
}
\end{center}

\bibliographystyle{ieeetr}
\bibliography{apssamp,bib/ml,bib/tool,bib/rl,bib/vqc,bib/qml_examples,bib/quantumRL,bib/evo,bib/tn,bib/tn_ml,bib/qc,bib/classical_evo_ml,bib/refs,bib/rl_applications}%

\end{document}